\documentclass[reprint,showpacs,
aps]{revtex4-1}

\usepackage{amsmath, amssymb}
\usepackage{type1cm}
\usepackage[dvipdfmx]{color}
\usepackage[dvipdfmx]{graphicx}
\usepackage{dcolumn}
\usepackage{bm}

\begin{document}

\preprint{APS/123-QED}

\title{Supersymmetry Breaking and Nambu-Goldstone Fermions in Interacting Majorana Chains} 

\author{Noriaki Sannomiya\thanks{E-mail address: sannomiya@cams.phys.s.u-tokyo.ac.jp} and Hosho Katsura}

\affiliation{
Department of Physics, Graduate School of Science, The University of Tokyo, Hongo, Tokyo 113-0033, Japan}

\date{\today}
\begin{abstract}
We introduce and study a lattice fermion model in one dimension with explicit \mbox{$\mathcal{N}=1$} supersymmetry (SUSY). The Hamiltonian of the model is defined by the square of a supercharge built from Majorana fermion operators. The model describes interacting Majorana fermions and its properties depend only on a single parameter $g$. When $g=1$, we find that SUSY is unbroken and the ground states are identical to those of the frustration-free Kitaev chains. We also find a parameter regime in which SUSY is restored in the infinite volume limit. For sufficiently large $g$, we prove that SUSY is spontaneously broken and the low-lying excitations are gapless, which can be thought of as Nambu-Goldstone fermions. 
We then show numerically that these gapless modes have cubic dispersion at long wavelength.

\end{abstract}

\pacs{71.10.Fd, 71.10.Pm, 11.30.Pb}
\maketitle



\section{\label{sec:level1}Introduction}
Spontaneous symmetry breaking (SSB) is one of the most important concepts in physics, and it is known that SSB gives rise to gapless excitations called Nambu-Goldstone~(NG) bosons~\cite{PR_Nambu,NC_Goldstone,PR_Goldstone}. In recent years, counting theories of NG bosons in non-relativistic systems have attracted much attention~\cite{PRL_Watanabe,PRL_Hidaka}. 
When the generators of broken symmetry are fermionic, it is expected that SSB leads to massless fermions instead of massless bosons at low energies. The most famous example of such fermionic symmetries is supersymmetry (SUSY) ~\cite{Wess_NPB74,Witten_NPB82}. 
SUSY is a symmetry that relates bosons and fermions, and is expected to solve some fundamental problems such as the hierarchy problem~\cite{PRD_Weinberg,PRD_Gildener}. Despite its importance, SUSY has yet to be confirmed experimentally. Thus, SUSY is considered to be spontaneously broken if realized in nature. It is known, in relativistic systems, that spontaneous SUSY breaking gives rise to massless excitations called NG fermions or Goldstinos~\cite{Salam:1974zb}. The low-energy properties of such systems can be described by the theory of non-linear realization of SUSY~\cite{PLB_Volkov}.

In condensed matter physics, a few examples of SUSY have been discussed in the context of lattice models~\cite{JPA_Nicolai_76,JPA_Nicolai_77,PRL_Fendley_2003,PRD_Fu}, cold atomic systems~\cite{PRL_Yu,PRA_Hidaka,PRA_Hidaka_2017}, and emergent SUSY at criticality~\cite{PRB_Lee_2007,Science_Grover,PRL_Jian_2015,PRL_Jian_2017,PRL_Rahmani, PRL_Huijse}.
The relation between spontaneous SUSY breaking and NG fermions in non-relativistic systems was also studied in the 
literature. In particular, it was argued that NG fermions with quadratic dispersion associated with spontaneous SUSY breaking can be realized in cold atomic systems with a mixture of bosons and fermions~\cite{PRL_Yu,PRA_Hidaka,PRA_Hidaka_2017}. On the other hand, in our previous work, we studied extensions of Nicolai's model~\cite{JPA_Nicolai_76,JPA_Nicolai_77} and found that spontaneous SUSY breaking gives rise to NG fermions with linear or cubic dispersion relation, depending on the details of the models~\cite{U1Nicolai,Z2Nicolai}. 
So far, these studies have been limited to $\mathcal{N}=2$ supersymmetric models. Thus, the properties of $\mathcal{N}=1$ SUSY in non-relativistic systems have not been elucidated.

The concept of Majorana fermions is also one of the most important ideas in high energy and condensed matter physics, and it has been attracting renewed attention in terms of the application to quantum information~\cite{RMP_Nayak}. To date, considerable effort has been devoted to the studies 
of 
free Majorana fermions and they are relatively well-understood. 
On the other hand, the effects of interactions on Majorana fermions remain elusive and have been the focus of recent research, as they can potentially lead to a variety of interesting phenomena such as the reduction of the topological classification of free fermions~\cite{PRB_Fidkowski_2010,PRB_Fidkowski_2011}.  
In addition, models of interacting Majorana fermions exhibit rich phase diagrams including emergent SUSY~\cite{PRL_Rahmani,PRB_Rahmani,PRB_Milsted,PRB_Zhu,PRB_Affleck}.

In this paper, we introduce and study 
a lattice fermion model with $\mathcal{N}=1$ SUSY, 
which describes a chain of interacting Majorana fermions. 
The Hamiltonian 
consists of quadratic and quartic terms in Majorana fermions and its properties depend only on a single parameter $g \in \mathbb{R}$. 
The model has an exact SUSY 
and allows us to study spontaneous SUSY breaking and NG fermions in non-relativistic situations. When $|g|=1$, SUSY is unbroken and the ground states can be obtained analytically. 
For $|g| < g_{\rm c} \approx 1$, we find that 
SUSY is restored in the infinite volume limit. 
On the other hand, for sufficiently large $|g|$, we show that spontaneous SUSY breaking takes place in both finite and the infinite chains.  
We then prove the existence of gapless excitations, which are the analogue of NG fermions in non-relativistic systems, by using a variational argument. 
We also show numerically that these gapless modes 
have cubic dispersion at long wavelength.

The structure of this paper is as follows. In Sec. \ref{sec:model}, we introduce the Hamiltonian as the square of a supercharge. In Sec. \ref{sec:break}, we first review the definition of spontaneous SUSY breaking. Then, we discuss conditions under which SUSY is unbroken, broken finite system but restored in the infinite system,  or broken in both finite and the infinite systems. In Sec. \ref{sec:NGf}, we prove the existence of NG fermions using elementary inequalities, and show that their dispersion relation is cubic. We conclude our paper in Sec. \ref{sec:conclusion}. In Appendix \ref{sec:JWHam}, we give an explicit expression for the Jordan-Wigner transformed Hamiltonian. In Appendixes \ref{sec:GSene}, \ref{sec:variation} and \ref{sec:fourier}, we derive formulas which are used in the main text. In Appendix \ref{sec:scaling}, we discuss the finite-size scaling of the ground-state energy for some parameters. In Appendix \ref{sec:traop}, we give an explicit representation of translation operator of Majorana fermion by one site.
\smallskip

\section{Model}
\label{sec:model}
In this section, we introduce $\mathcal{N}=1$ supersymmetric lattice fermion model on $(1+1)$-dimensional lattice.
We consider the following supercharge
\begin{align}
Q=\sum_{j=1}^Ng\gamma_j+\mathrm{i}\gamma_{j-1}\gamma_j\gamma_{j+1},
\label{eq:suQ}
\end{align}
where $\gamma_j=\gamma_j^\dagger$ is a Majorana fermion operator acting on $j$-th site which satisfies the Clifford algebra \mbox{$\{\gamma_i,\gamma_j\}=2\delta_{i,j}$}, 
$g$ is a real parameter, and $N~(\ge4)$ is the (even) number of sites. We assume the periodic boundary conditions (PBC), i.e., $\gamma_{j+N}=\gamma_j$.  
The supercharge is hermitian ($Q^\dagger=Q$) and is invariant under the translation by one Majorana site $T$~:~$\gamma_j\to\gamma_{j+1}$. The supercharge is fermionic since it anticommutes with the fermionic parity $(-1)^F:= \mathrm{i}^{N/2}\prod_{j=1}^N\gamma_{j}$. 
The Hamiltonian is defined 
as $H=Q^2$. The supercharge $Q$ is conserved quantity by definition, and thus the model has an explicit $\mathcal{N}=1$ SUSY. The Hamiltonian is positive semi-definite since the expectation value in arbitrary state $|\psi\rangle$ is non-negative, i.e., \mbox{$\langle\psi|H|\psi\rangle$}$=$\mbox{$\parallel Q|\psi\rangle\parallel^2\ge0$}. Without loss of generality, we can assume that the parameter $g$ is non-negative since 
the model with $-g$ can be mapped to the one with $g$ by 
sending $\gamma_j\to\gamma_{N-j}$.

\begin{figure}[htb]
\includegraphics[width=0.95\columnwidth]{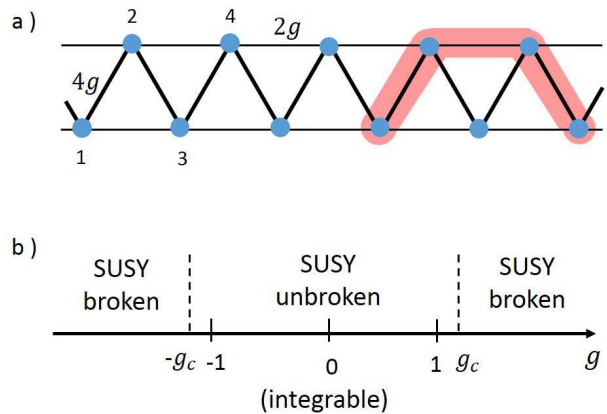}
\caption{(a) Schematics of the Hamiltonian. Symbols $4g$ and $2g$ represent the first and second terms of the free Hamiltonian [Eq.~(\ref{eq:fHam})], respectively. The pink region represents 
a quartic interaction of Majorana fermions described by Eq.~(\ref{eq:iHam}). (b) Schematic phase diagram of the Hamiltonian~(\ref{eq:Ham}) in the infinite volume limit as a function of $g$. At $g=\pm 1$, the ground states can be obtained analytically. The model at $g=0$ is another solvable case~\cite{Fendley}. 
The SUSY broken/unbroken transition occurs at $g=\pm g_{\rm c}$ with $g_{\rm c}$ slightly larger than $1$. }
\label{fig:schema}
\end{figure}

The Hamiltonian can be written more explicitly as
\begin{align}
H=H_{\rm free}+H_{\rm int}+Ng^2
\label{eq:Ham}
\end{align}
with
\begin{align}
H_{\rm free}= & 2g \mathrm{i}\sum_{j=1}^N(2\gamma_j\gamma_{j+1}-\gamma_{j-1}\gamma_{j+1}), \label{eq:fHam} \\
H_{\rm int}= & \sum_{j=1}^N(1-2\gamma_{j-1}\gamma_j\gamma_{j+2}\gamma_{j+3}).\label{eq:iHam} 
\end{align}
The first term $H_{\rm free}$ describes the hopping of Majorana fermions between nearest- or next-nearest-neighbor sites, while the second term $H_{\rm int}$ describes quartic interactions (See also Fig.~\ref{fig:schema} (a)). 
The second term in Eq.~(\ref{eq:fHam}) is not invariant under the time-reversal operation ${\cal K}$~:~$\gamma_j\to(-1)^{j+1}\gamma_j$, $\mathrm{i}\to-\mathrm{i}$. Note that each summand in Eq. (\ref{eq:iHam}) is positive semi-definite.  

\smallskip

\section{SUSY Breaking}
\label{sec:break}
In this section, we first give a definition of SUSY breaking, and then we discuss the property of the ground states in terms of SUSY breaking. In Sec. \ref{sec:unbroken}, we show that SUSY is unbroken for $g=1$, and ground state can be calculated analytically. In Sec. \ref{sec:restoration}, we find points where SUSY is broken in finite systems but restored in the infinite volume limit. In Sec. \ref{sec:breaking}, we prove that SUSY is broken when $g$ is larger than $8/\pi$ by deriving a lower bound on the ground-state-energy density. 
\subsection{Definition of SUSY breaking}
\label{sec:defsusy}
We now review the precise definition of spontaneous SUSY breaking~\cite{U1Nicolai}:

\smallskip

\noindent
{\bf Definition:}~{\it SUSY is said to be spontaneously broken if the ground-state-energy density is positive.}

\smallskip

This definition is well-defined in both finite and infinite systems. In the large-$g$ limit, SUSY is spontaneously broken in our model since the constant term in Eq.~(\ref{eq:Ham}) proportional $g^2$ is dominant. In the rest of this paper, we 
mostly focus on spontaneous SUSY breaking for modest values of $g$.

\smallskip

\subsection{SUSY unbroken case}
\label{sec:unbroken}
In this subsection, we discuss the properties of the ground states for $g=1$. In this case, SUSY is unbroken, i.e., there exist zero-energy states. These states must be ground states since the Hamiltonian is positive semi-definite. Although the system is interacting, ground states can be obtained analytically in this case. One of the ground states is identical to the trivial ground state of the Kitaev chain~\cite{Kitaev_chain} ($t=\Delta=0$,~$\mu<0$) with PBC, and the other is the same as the topological ground state of the Kitaev chain ($t=\Delta\neq0$,~$\mu=0$) with PBC. We refer to these two states as $|\Psi_0\rangle$ and $|\Psi_1\rangle$, respectively. 
They are related to each other by translation operator $T$. We note that they are eigenstates of the fermionic parity $(-1)^F$ with opposite eigenvalues, as $T$ anticommutes with $(-1)^F$~\cite{PRL_Hsieh}.

In order to verify that these two states are ground states of our Hamiltonian, we rewrite 
$Q$ in Eq. (\ref{eq:suQ}) as
\begin{align}
Q & =\sum_{l=1}^{L}(\gamma_{2l-2}+\gamma_{2l+1})(1+\mathrm{i}\gamma_{2l-1}\gamma_{2l}), \\
& =\sum_{l=1}^{L}(\gamma_{2l-1}+\gamma_{2l+2})(1+\mathrm{i}\gamma_{2l}\gamma_{2l+1}),
\end{align}
where $L=N/2$. The two states  $|\Psi_0\rangle$ and $|\Psi_1\rangle$ are, respectively, annihilated by the operators $(1+\mathrm{i}\gamma_{2l-1}\gamma_{2l})$ and $(1+\mathrm{i}\gamma_{2l}\gamma_{2l+1})$ for all $l$, i.e., 
\begin{align}
(1+\mathrm{i}\gamma_{2l-1}\gamma_{2l})|\Psi_0\rangle =0, 
\quad
(1+\mathrm{i}\gamma_{2l}\gamma_{2l+1})|\Psi_1\rangle =0 \label{eq:topo}.
\end{align}
From this it follows that $Q|\Psi_0\rangle=Q|\Psi_1\rangle=0$. Note that although the Hamiltonian breaks time reversal symmetry explicitly, each ground state is invariant under the time reversal operation~$\mathcal{K}$.

We find numerically that the lowest excitation energy decays exponentially with increasing the system size $N$. Therefore, the lowest excited states are expected to be degenerate with the ground states in the infinite volume limit. 
For finite systems, we have checked numerically that the number of the ground states is two unless $N \equiv 0~({\rm mod}~8)$. When $N$ is a multiple of $8$, we have four ground states, 
two of which are $|\Psi_0\rangle$ and $|\Psi_1\rangle$ as discussed above. 
The other two can also be obtained analytically. The explicit expression for one of the rest ground states is 
\begin{align}
|\Phi_0\rangle=\frac{1}{N_0}\sum_{j=1}^Ne^{-\mathrm{i}\frac{\pi}{4}j}\gamma_j|\Psi_0\rangle,
\end{align}
where $N_0$ is a normalization factor. The coefficient $e^{-\mathrm{i}\frac{\pi}{4}j}$ indicates 
why this state is a ground state when $N$ is a multiple of $8$.
Due to the translational symmetry, the other ground state can be obtained by acting with $T$ on $|\Phi_0\rangle$.

\smallskip

\subsection{SUSY restoration case}
\label{sec:restoration}
\begin{figure}[htb]
\includegraphics[width=0.9\columnwidth]{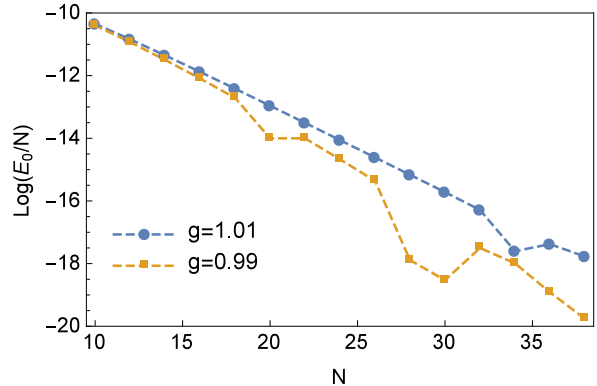}
\caption{Log-linear plots of ground-state-energy density for $g=0.99$ and $1.01$ as a function of $N$. Here, $E_0$ refers to the ground-state energy.}
\label{fig:gsed}
\end{figure}

For $g\neq1$, we have verified that SUSY is broken spontaneously in finite systems by exact numerical diagonalization. 
However, when the parameter $g$ is close to $1$, we find that the ground-state-energy density tends to decrease exponentially with the system size. 
In Fig.~\ref{fig:gsed}, we show the log-linear plot of the ground-state-energy density as a function of $N$ for $g=0.99,~1.01$ and $N=10, \dots, 38$. 
The results suggest that SUSY is restored in the infinite volume limit, although it is broken in a finite volume. 

\begin{figure}[htb]
\includegraphics[width=0.9\columnwidth]{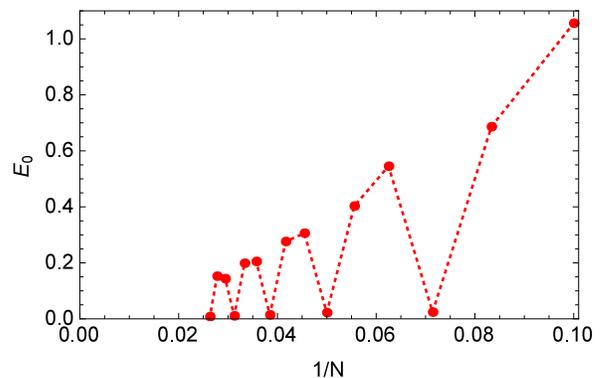}
\caption{The ground-state energy $E_0$ for $g=0$ as a function of $1/N$.}
\label{fig:gse0}
\end{figure}

We also find that SUSY is restored in the infinite volume limit even when $g$ is not close to $1$. 
In Fig.~\ref{fig:gse0}, we plot the ground-state energy $E_0$ as a function of $1/N$ for $g=0$ and $N=10, \dots, 38$.
\noindent
From Fig.~\ref{fig:gse0}, we see that the ground-state energy has a periodic structure depending on the number of sites $N$ (mod $6$)~\footnote[3]{A similar periodic structure depending on $N$ (mod $8$) 
was found in the $\mathcal{N}=1$ supersymmetric Sachdev-Ye-Kitaev model in Ref.~\cite{PRD_Fu}}. Since the Hamiltonian is described by the sum of local operators, 
one may expect that the ground-state energy is of the order of $N$. However, Fig.~\ref{fig:gse0} implies that the ground-state energy converges to a finite value in the infinite volume limit, so that the ground-state-energy density goes to zero. Thus, SUSY is restored in the infinite volume limit.

We note in passing that the model with $g=0$ is integrable and the SUSY restoration can be proved analytically~\cite{Fendley}.
The restoration of SUSY is not observed in our previous $\mathcal{N}=2$ SUSY models where SUSY is unbroken in the point $g=0$.
Since SUSY is unbroken for both $g \approx 1$ and $g=0$, we expect that there is an extended region of the phase diagram in which SUSY is restored 
(see Fig.~\ref{fig:schema} (b)). 
However, the precise location of the phase boundaries ($\pm g_{\rm c}$) and the nature of the phase transition remain unclear and require further study~\cite{PRL_OBrien}.

\smallskip

\subsection{SUSY breaking case}
\label{sec:breaking}
Next, we prove that SUSY is spontaneously broken for $g > 8/\pi$ in both finite and infinite systems. The proof goes as follows. The ground-state energy of the free part $E_0^{\rm free}$ reads,
\begin{align}
E_0^{\rm free}=-\frac{8g}{\tan(\pi/N)},
\label{eq:freeE}
\end{align}
(see Appendix \ref{sec:GSene} for details). Using Anderson's argument~\cite{PR_Anderson,PRB_Valent,PRD_Beccaria,PRL_Nie}, we get a lower bound for the ground-state-energy density $E_0/N$ as follows:
\begin{align}
\frac{E_0}{N} \ge g^2+\frac{E_0^{\rm free}}{N}\ge g\left(g-\frac{8}{\pi}\right).
\label{eq:LB}
\end{align}
Here, for the first inequality, we have used the fact that $H_{\rm int}$ in Eq.~(\ref{eq:Ham}) is positive semi-definite. 
Clearly, the inequality Eq.~(\ref{eq:LB}) shows that SUSY is spontaneously broken when $g>8/\pi$. This condition is not optimal since numerical results show that the ground-state-energy density is positive in both finite and the infinite systems even when $g$ is smaller than $8/\pi$ as shown in Appendix \ref{sec:scaling}.

\smallskip

\section{Nambu-Goldstone fermions}
\label{sec:NGf}
In the previous section, we have shown that SUSY is broken when $g$ is larger than $8/\pi$. In this section, we prove the existence of NG fermions, and show that NG fermions have cubic dispersion relation. In Sec. \ref{sec:existence}, we prove that low-energy excitation states are bounded from above by $p$-linear when $g > 8/\pi$. In Sec. \ref{sec:dispersion}, we clarify that the dispersion relation of NG fermion is cubic in momentum by employing both analytical and numerical methods.
\subsection{Existence of Nambu-Goldstone fermions}
\label{sec:existence}
In this subsection, we give a proof of the existence of the gapless fermionic excitations associated with spontaneous SUSY breaking using a variational method based on the Bijl-Feynman ansatz~\cite{PR_Feynman,ZPB_Horsch,PRB_Stringari,JPSJ_Momoi}. 
Suppose that $g>8/\pi$, so that SUSY is spontaneously broken. 
Take a variational state $|\psi(p)\rangle=Q_p|\psi_0\rangle$, where $|\psi_0\rangle$ is a SUSY broken ground state and 
\begin{align}
Q_p=\sum_{j=1}^Nq_j\cos(pj) ,
\end{align}
is the Fourier component of the local supercharge $q_j = g\gamma_j+\mathrm{i}\gamma_{j-1}\gamma_j\gamma_{j+1}$. 
Here, the momentum $p$ takes the values of $2\pi m/N$ ($m\in \mathbb{Z}$). The definition of $Q_p$ immediately implies that it is Hermitian ($Q_p^\dagger=Q_p$) and is an even function of $p$ ($Q_{-p}=Q_p$). 
We note that the ground states are at least doubly degenerate in this case and 
another ground state is $Q|\psi_0\rangle$. 
For $p \ne 0$ (${\rm mod}~2\pi$), the variational state $|\psi(p)\rangle$ is orthogonal to both $|\psi_0\rangle$ and $Q|\psi_0\rangle$ since $|\psi(p)\rangle$ can be written as a linear combination of states with momenta $\pm p$.

We define a variational energy in terms of the trial state $|\psi(p)\rangle$ as 
\begin{align}
\epsilon_{\rm var}(p)=\frac{\langle\psi(p)|H|\psi(p)\rangle}{\langle\psi(p)|\psi(p)\rangle}-E_0.
\end{align}
Here, $E_0$ is the ground-state energy of the Hamiltonian. This variational energy $\epsilon_{\rm var}(p)$ is larger than or equal to the first excitation energy. Using the Pitaevskii-Stringari inequality~\cite{JLTP_Pitaevskii} and properties of the operator $Q_p$, one finds that, when $p$ is small enough, the variational energy is bounded by $p$ linear from above as shown in Appendix \ref{sec:variation},
\begin{align}
\epsilon_{\rm var}(p)\le\sqrt{\frac{C}{2E_0/N}}|p|+O(p^3).
\label{eq:NGf}
\end{align}
Here, $C$ is a constant of 
$O(1)$. From Eq.~(\ref{eq:NGf}), we 
see that there exist gapless excitations associated with spontaneous SUSY breaking. We note that these excitations are considered to be NG fermions since the trial sate $|\psi(p)\rangle$ has an opposite fermionic parity from that of the ground state $|\psi_0\rangle$.

\smallskip

\subsection{Dispersion relation}
\label{sec:dispersion}
Next, we discuss the dispersion relation of 
low-lying excitations. In the large-$g$ limit, $H_{\rm free}$ in Eq.~(\ref{eq:Ham}) is dominant. Thus, let us consider only this term for the moment. The Fourier transform of the free Hamiltonian $H_{\rm free}$ can be written as follows:
\begin{align}
H_{\rm free}=8g\sum_{p>0}f(p)\gamma^\dagger(p)\gamma(p)-\frac{8g}{\tan(\pi/N)}.
\label{eq:fourier}
\end{align}
Here, $\gamma(p)$ is the Fourier transform of local Majorana operators, and $f(p)$ is defined as $f(p):=2\sin(p)-\sin(2p)$. Details of calculation are shown in Appendix \ref{sec:fourier} In this case, the constant term in Eq.~(\ref{eq:fourier}) coincides with the ground-state energy. When the momentum $p$ is small enough, the dispersion relation of $H_{\rm free}$ is cubic since $f(p)\propto p^3$ around the origin.

\begin{figure}[htb]
\includegraphics[width=0.9\columnwidth]{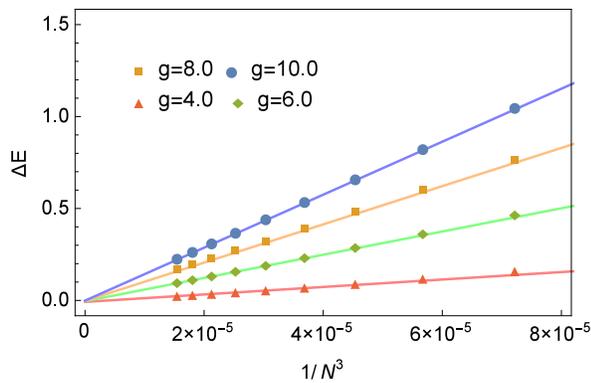}
\caption{The first excitation energy of the Hamiltonian for each $g$ as a function of $N^3$. 
Here, the number of sites $N$ is varied from $24$ up to $40$. The lines are fits to the data of $N=32, \dots,40$.}
\label{fig:1st}
\end{figure}

To see the dispersion for finite $g$, we calculate the many-body spectrum using exact diagonalization method with the help of translation operator of Majorana fermions discussed in Appendix \ref{sec:traop}. Figure \ref{fig:disp} shows the results for $g=8$ with $N=16, \dots, 24$. 
The excitation energies $\epsilon(p)$ are plotted with respect to the momentum $p$. 
Here, we redefine the momentum with the ground state being a zero-momentum state. The 
dotted curve is spectrum of the free part of the Hamiltonian described by $8gf(p)$. In the vicinity of 
$p=0$, there are energy levels that fit the curve. This implies a cubic dispersion at low energies. 
\begin{figure}[htb]
\includegraphics[width=0.9\columnwidth]{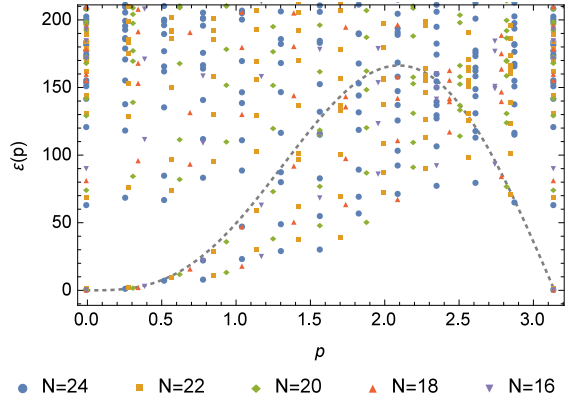}
\caption{Many-body spectrum of the Hamiltonian with $g=8$ as a function of momentum $p$. The dotted curve is one particle spectrum of $H_{\rm free}$ described by $8gf(p)$ (see Eq.~(\ref{eq:fourier})).}
\label{fig:disp}
\end{figure}
In order to verify this, we calculate the first excitation energy $\Delta E$ using exact diagonalization  
up to $N=40$ Majorana sites. The results are shown in Fig. \ref{fig:1st}. 
We plot the first excitation energy for $g=4,\dots,10$ as a function of $1/N^3$ and the lines are fits to the data of $N=32, \dots, 40$. 
The results clearly show that the lowest-excitation energy is proportional to $p^3$. 
Therefore, we conclude that the low-lying excitations have cubic dispersion when SUSY is spontaneously broken.

\smallskip

\section{Conclusion}
\label{sec:conclusion} 
In summary, we have introduced and studied 
a lattice fermion model in one dimension with $\mathcal{N}=1$ SUSY. The Hamiltonian is defined by the square of a supercharge $Q$ made up solely of Majorana fermions and depends only on the parameter $g$. At $|g|=1$, SUSY is unbroken and the ground states are identical to the ground states of the frustration-free Kitaev chains. While $\mathcal{N}=2$ SUSY is unbroken only at the point $g=0$ in our previous works, we found that $\mathcal{N}=1$ SUSY is broken spontaneously in finite systems yet restored in the infinite volume limit for $|g|<g_{\rm c} \approx 1$.
For $|g|>8/\pi$, we proved that SUSY is broken spontaneously and there exist gapless excitations which can be thought of as Nambu-Goldstone fermions. Using numerical methods, we showed that the lowest-excited states have cubic dispersion at long wavelength. We expect that our results provide a first step towards a comprehensive understanding of spontaneous SUSY breaking in both relativistic and non-relativistic models of interacting Majorana fermions.

\bigskip

\begin{acknowledgments}
The authors would like to thank P. Fendley for valuable comments. This work was supported, in part, by JSPS KAKENHI Grants \mbox{No. JP15K17719}, \mbox{No. JP16H00985}, \mbox{No. JP18K03445} and \mbox{No. JP18H04478}.
\end{acknowledgments}

\appendix

\section{Jordan-Wigner transform of the Hamiltonian}
\label{sec:JWHam}
In this section, we map the Hamiltonian to that of spin system with $S=1/2$ using the Jordan-Wigner transformation: 
\begin{align*}
\gamma_{2j-1}=\sigma_j^x\prod_{k=1}^{j-1}(-\sigma_k^z)\quad,\quad\gamma_{2j}=-\sigma_j^y\prod_{k=1}^{j-1}(-\sigma_k^z).
\end{align*}
The Hamiltonian in terms of spin operators reads
\begin{widetext}
\begin{align*}
H = & (g^2+1)N+2g\sum_{l=1}^{N/2}\left\{2(\sigma_l^z-\sigma_l^x\sigma_{l+1}^x) -\left(\sigma_l^x\sigma_{l+1}^y-\sigma_l^y\sigma_{l+1}^x\right)\right\} \\
 & -2\sum_{l=1}^{N/2}(\sigma_l^z\sigma_{l+1}^x\sigma_{l+2}^x+\sigma_l^x\sigma_{l+1}^x\sigma_{l+2}^z)\\
 & +((-1)^F +1)\left\{4g\sigma^y_{N/2}\sigma^y_1+2g(\sigma_{N/2}^x\sigma_1^y - \sigma^y_{N/2}\sigma_1^x)+2(\sigma_{N/2-1}^z\sigma_{N/2}^y\sigma_1^y + \sigma_{N/2}^y\sigma^y_1\sigma_2^z)\right\}.
\end{align*}
\end{widetext}
Since we assume periodic boundary conditions in terms of fermion operators, the Hamiltonian in terms of spin operators contains boundary terms which are proportional to $(-1)^F +1$. We note that this Hamiltonian contains similar terms which appear in \cite{PRL_OBrien}.

\section{The ground state energy of $H_{\rm free}$}
\label{sec:GSene}
In this section, we calculate the ground-state energy of the free part of the 
Hamiltonian $H_{\rm free}$ for finite length $N$.
The ground-state energy of $H_{\rm free}$ can be calculated by methods used in Ref~\cite{Kitaev_chain}.
\begin{align*}
H_{\rm free} & =  2\mathrm{i} g \sum_{j=1}^{N}(2\gamma_j\gamma_{j+1}-\gamma_{j-1}\gamma_{j+1}) = \frac{\mathrm{i}}{4}\Gamma^t A \Gamma,
\end{align*}
Here, $\Gamma$ is a vector of the form $(\gamma_1, \gamma_2,\dots, \gamma_N)^t$ and $A$ is the following $N\times N$ real skew-symmetric matrix 
\begin{align*}
A=g\begin{pmatrix}
~0~ & 4 & -2 & 0 & 0 &  \dots & 0 & 0 & 2 & ~-4~ \\
~-4~ & 0 & 4 & -2 & 0 &  \dots & 0 & 0 & 0 & ~2~ \\
2 & -4 & 0 & 4 & -2 &  \dots & 0 & 0 & 0 & ~0~ \\
\vdots &   &   & \ddots &  &  & \ddots &  &  & ~\vdots~ \\
0 & 0 & 0 & 0 & 0 &  \dots & -4 & 0 & 4 & ~-2~ \\
-2 & 0 & 0 & 0 & 0 &  \dots & 2 & -4 & 0 & ~4~ \\
4 & -2 & 0 & 0 & 0 &  \dots & 0 & 2 & -4 & ~0~ 
\end{pmatrix}.
\end{align*}
Since $A$ is real skew-symmetric, $A$ can be block diagonalized using an orthogonal matrix $Q$,
\begin{align}
Q^TAQ=\bigoplus_{l=1}^{N/2}\begin{pmatrix}
~0~ & ~\epsilon_l~ \\
~-\epsilon_l~ & ~0~
\end{pmatrix}.
\end{align}
Here, $\pm\mathrm{i}\epsilon_l$ are eigenvalues of the matrix $A$. By analogy with the Kitaev chain~\cite{Kitaev_chain},
 the ground state energy can be calculated as
\begin{align*}
E_0^{\rm free}=-\frac{1}{2}\sum_{l=1}^{N/2}\epsilon_l.
\end{align*}
Next, we consider the following eigenvalue problem in order to get the eigenvalues,
\begin{align}
\tilde{A}\bm{v}=\epsilon\bm{v}.
\label{eq:eigen}
\end{align}
Here, we define an $N\times N$ Hermitian matrix $\tilde{A}$ as $\tilde{A}=\mathrm{i}A$ so that the eigenvalues are real.  
From the eigenvalue equation Eq.~(\ref{eq:eigen}), we obtain
\begin{align*}
2g\mathrm{i}(v_{j-2}-2v_{j-1}+2v_{j+1}-v_{j+2}) & =\epsilon v_{j}
\end{align*}
for each component $v_j \ (j=1,\dots, N \quad {\rm mod} \ N)$. 
Next, we assume the following ansatz
\begin{align*}
v_j=\alpha e^{\mathrm{i}pj},
\end{align*}
from which we get $e^{\mathrm{i}pN}=1$ for periodic boundary conditions $v_{j+N}=v_j$. From this, for all $j=1, \dots, N$, we obtain
\begin{align*}
\epsilon & =g2\mathrm{i}(e^{\mathrm{i}p(j-2)}-2e^{\mathrm{i}p(j-1)}+2e^{\mathrm{i}p(j+1)}-e^{\mathrm{i}p(j+2)})e^{-\mathrm{i}pj} \\
& =2g\mathrm{i}(2e^{\mathrm{i}p}-2e^{-\mathrm{i}p}-e^{2\mathrm{i}p}+e^{-2\mathrm{i}p}) \\ 
& =-8g\sin(p)+4g\sin(2p).
\end{align*}
Here, the momentum $p$ is an element of the set $\mathcal{M}$, which is defined as
\begin{align*}
\mathcal{M}=\left\{ 0,\pm\frac{2\pi}{N},\pm\frac{4\pi}{N},\dots,\pm\frac{(N-2)\pi}{N},\pi \right\}.
\end{align*}
The eigenvalues of the matrix $A$ are $\mathrm{i}(8g\sin(p)-4g\sin(2p))$ with $p\in \mathcal{M}$. 
Therefore, the ground-state energy $E_0^{\rm free}$ of $H_{\rm free}$ is obtained as
\begin{align*}
E_0^{\rm free} & =-\cfrac12 \sum_{l=1}^{N/2}\left(8g\sin\left(\frac{2 \pi l }{N}\right)-4g\sin\left(\frac{4\pi l}{N}\right)\right) \nonumber \\
 & = -\frac{8g}{\tan\left(\pi/N\right)}.
\end{align*}


\section{Finite size scaling of the ground-state energy density for $g\le8/\pi$}
\label{sec:scaling}
\begin{figure}[htb]
\includegraphics[width=1.0\columnwidth]{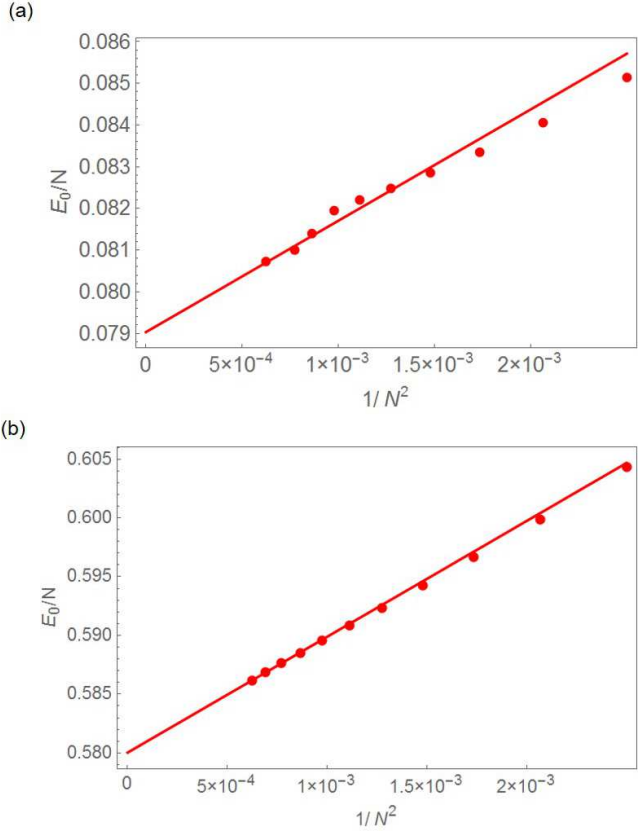}
\caption{The ground-state-energy density for $g=1.5$~(a) and $g=2$~(b) as a function of $1/N^2$. Lines are fits to the data of $N=34, \dots,40$. Estimated values of energy density in the thermodynamic limit are $0.079$ and $0.58\dots$, respectively.}
\label{fig:eden}
\end{figure}

We have proved in the main text that SUSY is spontaneously broken in both finite and the infinite systems when $g>8/\pi~(=2.546479\dots)$. However, this value is not optimal, and there is a region of the parameter $g$ in which SUSY is spontaneously broken even when $g\le8/\pi$. In order to verify this, we calculate the ground-state-energy density numerically. The results for $g=1.5$ and $g=2$ are shown in Fig.~\ref{fig:eden}.

In these figures, lines are fits to the data of $N=34,\dots, 40$. From these figures, we find that the ground-state-energy densities for $g=1.5$ and $g=2$ in the thermodynamic limit are $0.079$ and $0.58\dots$, respectively. Therefore, we conclude that there is a region of the parameter $g < 8/\pi$ in which SUSY is broken spontaneously.


\section{Variational method based on Bijl-Feynman ansatz}
\label{sec:variation}
In this section, we calculate the upper bound of the variational energy in order to prove that there exist gapless modes associated with SUSY breaking. By a straightforward calculation, we can rewrite the variational energy in terms of the double commutator of $H$ and $Q_p$,
\begin{align}
\epsilon_{\rm var}(p)=\frac{\langle[Q_p,[H,Q_p]]\rangle_0}{\langle\{Q_p,Q_p\}\rangle_0}.
\label{eq:variene}
\end{align}
Here, the symbol $\langle\cdots\rangle_0$ denotes the expectation value in the ground state. The local supercharge $q_j$ satisfies the following locality condition
\begin{align*}
\{q_i,q_j\}=\left\{ \begin{array}{ll}
{\rm non zero} & |i-j|\le 2 \\
0 & {\rm others} \\
\end{array} \right .
\end{align*}
Therefore, the commutator $[H,Q_p]$ can be written in terms of a sum of local operators. However, the double commutator $[Q_p,[H,Q_p]]$ may not so. In order to get an upper bound on $\epsilon_{\rm var}(p)$, we apply the Pitaevskii-Stringari inequality~\cite{JLTP_Pitaevskii}. By applying it to Eq. (\ref{eq:variene}), we obtain 
\begin{align*}
\epsilon_{\rm var}^2(p)\le\frac{\langle\{[H,Q_p],[Q_p,H]\rangle_0}{\langle\{Q_p,Q_p\}\rangle_0}=\frac{f_{\rm n}(p)}{f_{\rm d}(p)},
\end{align*}
where, $f_{\rm n}(p)$ and $f_{\rm d}(p)$ are defined by $\langle\{[H,Q_p],[Q_p,H]\rangle_0$ and $\langle\{Q_p,Q_p\}\rangle_0$, respectively. Since $f_{\rm n}(p)$ can be written as a sum of expectation values of local operators, it is of order of $N$. When $p=0$, we get $f_{\rm n}(0)=0$ and $f_{\rm d}(0)=2E_0$, respectively. Thus, when $p$ is small enough, we have $f_{\rm n}(p)=N(Cp^2+\mathrm{O}(p^4))$ and $f_{\rm d}(0)=2E_0+\mathrm{O}(p^2)$. Here, we use the facts that $Q_p$ is an even function of $p$~($Q_{-p}=Q_p$), and a Hermitian operator~($Q^\dagger_{p}=Q_p$). From these results, the variational energy is bounded by $p$-linear from above
\begin{align*}
\epsilon_{\rm var}(p)\le\sqrt{\frac{C}{2E_0/N}}|p|+\mathrm{O}(p^3).
\end{align*}
This proves that there exist gapless excitations associated with spontaneous SUSY breaking. We note that this gapless modes are considered to be NG fermion since the trial sate $|\psi(p)\rangle$ has a different fermionic parity from that of the ground state $|\psi_0\rangle$.


\section{Fourier transform of $H_{\rm free}$}
\label{sec:fourier}
In this section, we study, using the Fourier transformation, the dispersion relation of $H_{\rm free}$ defined by
\begin{align*}
H_{\rm free}=2g\mathrm{i}\sum_{j=1}^N(2\gamma_j\gamma_{j+1}-\gamma_{j-1}\gamma_{j+1}).
\end{align*}
Here, we assume PBC and 
$N$ even. The Fourier transform of Majorana fermion operators are defined as~\cite{PRB_Rahmani}
\begin{align*}
\gamma_j=\sqrt{\frac{2}{N}}\sum_{p}\gamma(p)e^{\mathrm{i}pj},
\end{align*}
and the Inverse Fourier transformation is also defined as
\begin{align*}
\gamma(p)=\sqrt{\frac{1}{2N}}\sum_{j=1}^N\gamma_je^{-\mathrm{i}pj}.
\end{align*}
Here, $p$ takes the values of $2\pi m/N$~($m\in \mathbb{Z}$) since PBC is assumed. From the Clifford algebra of Majorana fermion operators, Fourier transformed Majorana operators satisfy the following anti-commutation relation
\begin{align*}
\{\gamma(p),\gamma(p')\}=\delta_{p,-p'}.
\end{align*}
This relation implies $\gamma^\dagger(p)=\gamma(-p)$. The free part of the Hamiltonian $H_{\rm free}$ can be rewritten as
\begin{align*}
H_{\rm free} & =2g\mathrm{i}\sum_{j=1}\left(2\gamma_j\gamma_{j+1}-\gamma_{j-1}\gamma_{j+1}\right) \nonumber \\
& =4g\sum_p\left(2\sin(p)-\sin(2p)\right)\gamma(-p)\gamma(p) \nonumber \\
& =\sum_{p>0}\left(2\sin(p)-\sin(2p)\right)\gamma(-p)\gamma(p)-8g\sum_{p>0}\sin(p).
\end{align*}
By a straightforward calculation, we obtain
\begin{align*}
H_{\rm free} =8g\sum_{p>0}(2\sin(p)-\sin(2p))\gamma^\dagger(p)\gamma(p)-\frac{8g}{\tan(\pi/N)}.
\end{align*}
Here, we note that the last constant term coincides with the ground state energy as calculated in Sec.~\ref{sec:GSene}.
From this, we see that the dispersion relation is cubic in momentum when the momentum $p$ is small enough, i.e.,
\begin{align*}
8g(2\sin(p)-\sin(2p))\sim 8g|p|^3.
\end{align*}


\section{Translation operator of Majorana chain}
\label{sec:traop}
The importance of the translation operator in lattice Majorana fermions was discussed in Ref.~\cite{PRL_Hsieh}.
However, an explicit expression for the operator was not given there. In this section, we provide an explicit expression for the operator of translation by one Majorana site. 
Using Majorana fermion operators, we define the following operator $T_1$,
\begin{align*}
T_1=\gamma_1S_{1}\cdots S_{N-1}.
\end{align*}
Here, $S_{j}$ is defined as follows:
\begin{align*}
S_{j}=\frac{1}{\sqrt 2}(1+\gamma_j\gamma_{j+1}).
\end{align*}
This operator exchanges Majorana fermions on $j$-th 
and $(j+1)$-th sites up to a phase factor. 
$T_1$ translates the Majorana fermion on $i$-th site to that on $(i+1)$-th site,
\begin{align*}
T_1\gamma_jT_1^{-1}=\gamma_{j+1}.
\end{align*}
If $T_1$ is the translation operator, $T_1^N$ must be the identity. However we numerically find that the operator $T_1^N$ is $\pm1$ depending on $N$,
\begin{align*}
T_1^N=\left\{ \begin{array}{ll}
+1 & (N/2=4, 5, 8, 9,\dots) \\
-1 & (N/2=2, 3, 6, 7,\dots)
\end{array} \right .
\end{align*}
In the case of $N/2=2, 3, 6, 7,\dots$, we introduce the new operator $T_2$ defined by $T_2=e^{\mathrm{i}\pi/N}\ T_1$. We numerically verify that the operator $T_2$ satisfies $T_2^N=1$ in the case of $N/2=2, 3, 6, 7,\dots$. Operators $T_1$ and $T_2$ defined above become a translation operator for $N/2=4, 5, 8, 9,\dots$ and $N/2=2, 3, 6, 7,\dots$, respectively. Now, we define a new operator $T$ as 
\begin{align*}
T:=\left\{ \begin{array}{ll}
T_1 & (N/2=4, 5, 8, 9,\dots) \\
T_2 & (N/2=2, 3, 6, 7,\dots)
\end{array} \right .
\end{align*}
This operator $T$ is the translation operator of lattice Majorana fermions for all $N$.

\nocite{*}

\bibliography{MajoTrans}

\bigskip

\end{document}